\documentstyle[multicol,psfig,aps, prl]{revtex}

\date{\today}
\draft
\tighten
\begin{document}
\def\href#1#2{#2}
\def\sqr#1#2{{\vcenter{\hrule height.3pt
      \hbox{\vrule width.3pt height#2pt  \kern#1pt
         \vrule width.3pt}  \hrule height.3pt}}}
\def\square{\mathchoice{\sqr67\,}{\sqr67\,}\sqr{3}{3.5}\sqr{3}{3.5}}
\def\today{\ifcase\month\or
  January\or February\or March\or April\or May\or June\or July\or
  August\or September\or October\or November\or December\fi
  \space\number\day, \number\year}

\def\Bbb{\bf}
\topmargin=0.001in


\newcommand{\ww}{\mbox{\tiny $\wedge$}}
\newcommand{\pp}{\partial}

\title{Localized Gravity in String Theory}
\author{Andreas Karch and Lisa Randall}

\address {\qquad \\ Center for Theoretical Physics \&
Department of Physics\\
Massachusetts Institute of Technology\\
77 Massachusetts Avene\\
Cambridge, MA 02139, USA
}

\maketitle

\begin{abstract}

We propose a string realization of the AdS$_4$ brane
in AdS$_5$ that is known to localize gravity. Our
theory is $M$ D5 branes in the near horizon geometry of $N$ D3 branes,
where $M$ and $N$ are appropriately tuned.

\end{abstract}

\begin{multicols} {2}

String theories are only consistent with many additional
dimensions. Ever since the notion of extra dimensions was
introduced, it was believed that consistency with observed four
dimensional gravity requires these dimensions are of finite
extent, so it has been held that a viable string theory background
requires the compactification of the additional dimensions.
  Gravity localization   provides an
alternative to this viewpoint;  it is only necessary that a bound
state graviton mode dominates over the Kaluza-Klein modes
associated with the additional dimensions over the length scales
for which we have observed four-dimensional gravity \cite{kr}. 
In \cite{RS} an
effective five-dimensional Lagrangian was given which exhibits
this phenomenon. However, it seemed difficult to realize this
possibility in string theory or even a supersymmetric
theory \cite{herman,cveticrs,duffrs,cvetic,kallosh,nunez}.
One of the major difficultities
in the attempted constructions was the attempt to eliminate the
infinite volume region of space; that is geometry had to satisfy a
finite \lq \lq volume" condition; alternatively, there were stringent
conditions on the regions of space far from the localizing brane,
including the boundary. Attempts to realize the warped geometry in
string theory dealt with this through explicit compactification
so the is manifestly four-dimensional at long distances \cite{herman}.
However, it was recently shown that the localization idea is far
more general, and does not require anything about the regions of
space far from the brane \cite{kr}. The important point
is that gravity needs only be localized in a region of
finite extent; the full space might truly reflect the higher
dimensional geometry. This is a dramatically different picture
from compactification, which requires global properties of the
space-time. It is also a much weaker requirement on the geometry
than the original model \cite{RS} seemed to indicate. Furthermore,
the supergravity solution is only known explicitly for a very
limited class of objects within string theory; it seems quite
credible that the required geometric features could be realized in
a string background. We initiate an exploration of this
possibility in this paper.

Our proposal is that a set-up with $N$ D3 branes (say along 0126)
intersecting $M$ D5 branes (say along 012345) over a common
3-dimensional worldvolume can lead to localized gravity in a
4-dimensional AdS space. The fact that this set-up preserves
supersymmetry (in fact 16 supersymmetries) guarantees stability.

A special property of the AdS brane is that it lends
itself to a very natural holographic interpretation \cite{kr,long}.
As shown by Cardy, $SO(3,2)$ is the subgroup of
the 4d conformal group $SO(4,2)$ that preserves a given boundary.
This means the symmetry of AdS$_4$ is preserved.

As we showed in another paper
\cite{long} (see also \cite{BP}) the AdS$_5$ $\times$ $S^5$ near horizon
geometry set-up by the $N$ D3 branes, the 5-branes span an AdS$_4$
$\times$ $S^2$ worldvolume. This is very easy to see already from the
way AdS$_5$ $\times$ $S^5$ arises as the D3 near horizon
geometry: the dictionary of AdS/CFT comes with an identification of
flat space embedding coordinates of the D3 brane and the near horizon
geometry.
When AdS$_5$ is written in Poincare patch coordinates, the 4 Minkowski
directions are the 4 worldvolume directions of the D3 (0126).
The transverse directions are written in spherical coordinates.
The radial direction of this flat transverse space
becomes the radial direction in AdS$_5$ and the sphere surrounding the
D3 brane is the $S^5$ of the near horizon geometry. The D5 brane, whose
defining
equation in the flat embedding space coordinates is
$$x_6=x_7=x_8=x_9=0$$
hence has an AdS$_4$ $\times$ $S^2$ worldvolume,
both of curvature radius $L$. The latter is given by $x_7=x_8=x_9=0$ and
the former by $x_6=0$.

In this paper, we show that we expect such a brane
to be a viable candidate for localized gravity.
One can already argue from the gravity
side why such a wrapped 5-brane in the near horizon geometry of
the D3 branes should describe localized gravity. Forgetting for
the moment the issue of stability, consider a probe brane inside
the near horizon geometry of a stack of $N$ D3-branes. The near
horizon geometry is AdS$_5$ $\times$ $S^5$. A 5-brane wrapped over
an $S^2$ inside the $S^5$ effectively becomes a 3-brane in
AdS$_5$. A 3-brane living in the AdS$_5$ creates a warped geometry
of the sort considered in \cite{kr,kogan} (see also
\cite{gorski}), that is an AdS$_4$ brane.
This is possible  because five of the dimensions in the near
horizon region have already been compactified, so a 3-brane is
adequate to provide a warped AdS geometry.

We now  argue that the corresponding near horizon solution of the
intersecting D3-D5 configuration should localize gravity along the
lines of \cite{kr,andre,matthew,kogan}. The reasons for this are:
\begin{itemize}
\item {\bf The conformal symmetry:}
As shown in \cite{long}, the geometry set-up by the intersecting
D3-D5 configuration has a dual CFT description in terms
of a CFT with boundary. This dual makes it clear, that
$SO(3,2)$ symmetry is preserved, so our space time has
an AdS$_4$ slicing.
\item {\bf Asymptotics:}
Far away along the 6 direction the geometry is still that of the
D3 branes alone. So the geometry asymptotes to AdS$_5$
$\times$ $S^5$ for $r$ going to $\pm \infty$, where $r$ is the
transverse coordinate appearing in the warp factor.
\item {\bf Discrete symmetry:} The set-up with
the intersecting branes has a discrete $Z_2$ symmetry, $x_6
\rightarrow - x_6$. Hence the warp factor will have to have the
same symmetry, $r \rightarrow -r$.
\item {\bf Positive tension:} Since the energy density on the
5brane is positive, the jump in extrinsic curvature has to be
reflected by a jump in the warp factor that has the standard ``up-down''
shape.
\end{itemize}

At the moment the
full localized supergravity solution is not known
even though the results from \cite{long} should help to construct one.
Instead, we consider the back-reaction
from a five-dimensional effective theory
point of view.
In the 5d
string frame the Einstein-Hilbert term is given by
$$ \frac{L^5}{g_s^2} R $$
while the tension of
$M$ D5 branes wrapping the $S^2$ of radius $L^2$  is given by
$$ 4 \pi \; T_{D5} M L^2 $$  where
we need to recall that $T_{D5} \sim \frac{1}{g_s}$. So the jump in
extrinsic curvature is proportional to
$$M g_s.$$ 
First note that for $g_s \rightarrow 0$, $g_s N$
fixed, but $g_s M \rightarrow 0$, the backreaction can be
neglected. It was in this limit that
the probe calculation of \cite{long} established
that the D5 brane worldvolume is AdS$_4$ $\times$ $S^2$.
Clearly we need to get away from this limit to get a
significant backreaction and a localized graviton.

\end{multicols}

\begin{figure}
   \centerline{\psfig{figure=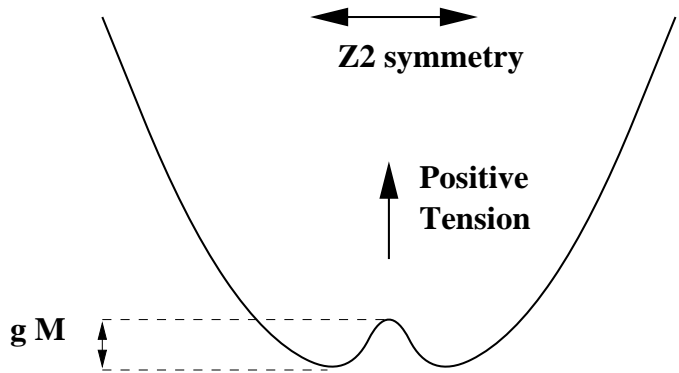,width=3.5in}}
    \caption{Turning on $g_s M$ at $q=0$.}
\label{localize}
 \end{figure}

\begin{figure}
   \centerline{\psfig{figure=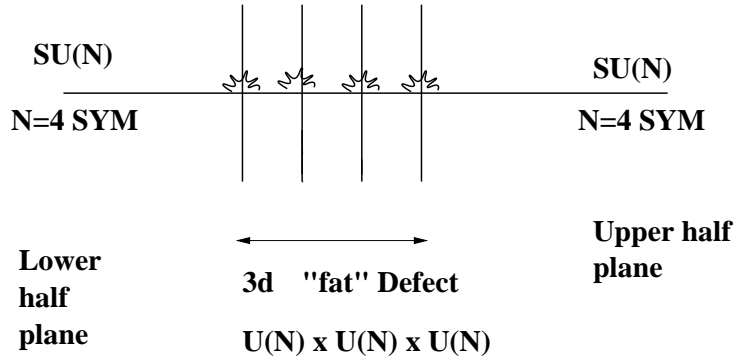,width=3.8in}}
    \caption{The fat domain wall givig rise to the CFT associated
with an $U(M)^3$ gauge theory for $M=4$ NS5 branes.}
\label{threedcft}
 \end{figure}

\begin{multicols}{2}

Fig. \ref{localize} summarizes the essential features of the
solution for  finite $g_s M$. As we increase the tension by
increasing $g_s M$,  $l$, the curvature
scale of the four-dimensional AdS space,   increases.
The critical value at which the 4d cosmological constant vanishes
in this approximation
 is
$$T_{crit}=\frac{3}{4 \pi G_N L}$$ where
$$G_N= (2 \pi)^3 g_s^2 l_s^8/ L^5$$ is the 5d Newton's
constant. Since $L\sim (g_s N)^{1/4}$, we see that the critical
tension is reached when $L^2 g_s M$ is of order $L^4 \sim g_s N$,
that is when $g_s M$ is of order $(g_s N)^{1/2}$. This is
the point at which the size of the 5-brane throat
$r_{throat} \sim (g_s M)^{1/2}$ (as computed
in flat space with an unwrapped 5-brane) becomes
comparable to the AdS$_5$ curvature radius $L$. We expect
that at this point  one can no longer
treat the wrapped 5-brane as a thin 3-brane brane in AdS$_5$. In order to
get an
analytic expression for $l$ in terms of $g_s M$ one
would need the corresponding gravity solution. We expect
as we approach the critical value, the effective field
theory approach where we neglect the effect of the 5-branes
on the sphere will break down, so one requires the full ten-dimensional
solution.
However, we anticipate
the essential properties of the solution will remain. It is reassuring that
the point at which a very flat brane happens, that is the
number of 5-branes is sufficient to generate
the critical tension, is when the analysis we have done
is just breaking down, so there should be a value
of $g_s M$ for which
$l>>L$. We therefore expect that the
set-up of intersecting D3 and D5 branes is a string theory
realization of localized gravity for sufficiently many 5-branes.

We can also argue for the viability of the theory
as a theory of localized gravity from the point of
view of the boundary CFT. 
The 5d gravitons give rise to the $\frac{1}{x^8}$
$TT$ correlators of the 4d CFT. A localized 4d mode
of the graviton corresponds to $\frac{1}{x^6}$
correlators on the boundary of the CFT.
This means from the dual point of view there should
exist a three dimensional field theory on
the boundary. The existence
of the three dimensional theory is a necessary (but
not sufficient) condition for gravity localization.
This was not satisfied, for example, in a theory
where $q$ $D3$ branes end on a single $D5$ brane.

In the language of the boundary CFT, for
$M \geq 2$ there exists an interacting 3d CFT on the boundary
whose degrees of freedom grow like $M N^2$.
To see this, replace for a moment the
D5 branes with NS5 branes and keep them at finite seperation. The
gauge theory realized by this brane configuration
is 4d $SU(N)$ with a ``fat'' defect which has a
3d $U(N)^{M-1}$ gauge theory with bifundamentals living on it,
the A-type quiver, as one can determine by
the usual rules \cite{HW}, see Fig.\ref{threedcft}. This
theory flows to an interacting CFT at the origin of its moduli space,
when we take the couplings to infinity, that is take the NS5 branes to
coincide. D5 branes give a mirror realization of the same CFT.
The stress energy
tensor of this 3d CFT  should couple to a mode of the 5d graviton localized
on the brane. This serves as a further check that
this theory is a reasonable candidate for gravity localization.

Now we can ask about going beyond the near horizon region. It is
important that localization is a local phenomenon; even if
asymptotically the three branes give flat space, one doesn't lose
the graviton and its associated four-dimensional gravitational
effects. All that happens is that in the regions where we don't
expect four-dimensional gravity, the KK contribution is not
suppressed relative to the graviton contribution. This is similar
to what happens for a positive tension brane in flat space; there
still is a graviton, but the KK modes are not suppressed since
there is no barrier in the volcano potential. What is important is
that there is {\it some} region with a barrier; the probe
calculation argues this is the case. This suffices for gravity for
length scales that are not too large.

The localized gravity set-up is still relatively young and we have
yet to realize the full range of possibilities. What is clear is
that one can localize gravity in such a way that only local
regions see the localized graviton as dominating the gravitational
force. This means one can have a set-up where four-dimensional
gravity only applies in specific regions of space. In some ways,
this is a rather compelling picture in that one doesn't need to
assume radical global properties of the full higher dimensional
space. What we have demonstrated in this paper is that
interesecting branes are sufficient to realize four-dimensional
gravity.

It is nice that this construction  involves multiple
branes, so that there is a realistic possibility that one can realize
gauge configurations, and in particular, the standard model, in
such a set-up.

Of course, the set-up as it now stands seems to rely on the fact
that we have AdS space, where current evidence is that we live in
dS space with small cosmological constant. However, it should be
borne in mind that the set-up as described is completely
superysymmetric. It is not clear how the geometry will be modified
with new sources of energy on the brane.

We have argued from an effective theory perspective that
we expect a range of $M$ where gravity localization is valid.
What is really required to verify this is the full supergravity
solution. The fact that this is difficult should not be
seen as an argument against localization. After all, we have
rigorously demonstrated the AdS nature of the brane, so
there must exist {\it some} supergravity solution. It
is an important and challenging problem to find it.

We conclude that localized gravity is most likely a viable
alternative to compactification. Because all curvature scales are
set by parameters of the theory (gauge charge and tension), many
of the difficult moduli problems should not be present. Of course,
there is a multiplicity of possible vacua corresponding to the
many possible brane set-ups. Perhaps it is hopeful that a
four-dimensional world can exist inside many possible brane
structures, since we are only sensitive to the local geometry.

\section*{Acknowledgements}
We would like to thank Andy Srominger for many useful suggestions
and observations. We would also like to thank Amanda Peet, Emil
Martinec, and Rob Myers, and Joe Polchinski.

\bibliography{3dcft}

\begingroup\raggedright\begin{thebibliography}{10}

\bibitem{kr}
A.~Karch and L.~Randall, ``Locally localized gravity,''
  \href{http://xxx.lanl.gov/abs/hep-th/0011156}{{\tt hep-th/0011156}}.

\bibitem{RS}
L.~Randall and R.~Sundrum, ``An alternative to compactification,'' {\em Phys.
  Rev. Lett.} {\bf 83} (1999) 4690--4693,
  \href{http://xxx.lanl.gov/abs/hep-th/9906064}{{\tt hep-th/9906064}}.

\bibitem{herman}
H.~Verlinde, ``Holography and compactification,'' {\em Nucl. Phys.} {\bf B580}
  (2000) 264--274, \href{http://xxx.lanl.gov/abs/hep-th/9906182}{{\tt
  hep-th/9906182}}.

\bibitem{cveticrs}
M.~Cvetic {\em et.~al.}, ``Randall-Sundrum brane tensions,''
  \href{http://xxx.lanl.gov/abs/hep-th/0011167}{{\tt hep-th/0011167}}.

\bibitem{duffrs}
M.~J. Duff, J.~T. Liu, and K.~S. Stelle, ``A supersymmetric type IIB
  Randall-Sundrum realization,''
  \href{http://xxx.lanl.gov/abs/hep-th/0007120}{{\tt hep-th/0007120}}.

\bibitem{cvetic}
K.~Behrndt and M.~Cvetic, ``Anti-de Sitter vacua of gauged supergravities with
  8 supercharges,'' {\em Phys. Rev.} {\bf D61} (2000) 101901,
  \href{http://xxx.lanl.gov/abs/hep-th/0001159}{{\tt hep-th/0001159}}.

\bibitem{kallosh}
R.~Kallosh and A.~Linde, ``Supersymmetry and the brane world,'' {\em JHEP} {\bf
  02} (2000) 005, \href{http://xxx.lanl.gov/abs/hep-th/0001071}{{\tt
  hep-th/0001071}}.

\bibitem{nunez}
J.~Maldacena and C.~Nunez, ``Supergravity description of field theories on
  curved manifolds and a no go theorem,''
  \href{http://xxx.lanl.gov/abs/hep-th/0007018}{{\tt hep-th/0007018}}.

\bibitem{long}
A.~Karch and L.~Randall, ``Open and closed string interpretation of SUSY CFT's
  on branes with boundaries,''
  \href{http://xxx.lanl.gov/abs/hep-th/0105132}{{\tt hep-th/0105132}}.

\bibitem{BP}
C.~Bachas and M.~Petropoulos, ``Anti-de-Sitter D-branes,''
  \href{http://xxx.lanl.gov/abs/hep-th/0012234}{{\tt hep-th/0012234}}.

\bibitem{kogan}
I.~I. Kogan, S.~Mouslopoulos, and A.~Papazoglou, ``A new bigravity model with
  exclusively positive branes,'' {\em Phys. Lett.} {\bf B501} (2001) 140--149,
  \href{http://xxx.lanl.gov/abs/hep-th/0011141}{{\tt hep-th/0011141}}.

\bibitem{gorski}
A.~Gorsky and K.~Selivanov, ``Tunneling into the Randall-Sundrum brane world,''
  {\em Phys. Lett.} {\bf B485} (2000) 271--277,
  \href{http://xxx.lanl.gov/abs/hep-th/0005066}{{\tt hep-th/0005066}}.

\bibitem{andre}
A.~Miemiec, ``A power law for the lowest eigenvalue in localized massive
  gravity,'' \href{http://xxx.lanl.gov/abs/hep-th/0011160}{{\tt
  hep-th/0011160}}.

\bibitem{matthew}
M.~D. Schwartz, ``The emergence of localized gravity,'' {\em Phys. Lett.} {\bf
  B502} (2001) 223--228, \href{http://xxx.lanl.gov/abs/hep-th/0011177}{{\tt
  hep-th/0011177}}.

\bibitem{HW}
A.~Hanany and E.~Witten, ``Type IIB superstrings, BPS monopoles, and
  three-dimensional gauge dynamics,'' {\em Nucl. Phys.} {\bf B492} (1997)
  152--190, \href{http://xxx.lanl.gov/abs/hep-th/9611230}{{\tt
  hep-th/9611230}}.

\end{thebibliography}\endgroup
\bibliographystyle{ssg}
\end{multicols}
\end{document}